\documentclass[fleqn,12pt,twoside]{article}
\usepackage{espcrc1}

\usepackage{graphicx}
\usepackage[figuresright]{rotating}

\newcommand{\AmS}{{\protect\the\textfont2
  A\kern-.1667em\lower.5ex\hbox{M}\kern-.125emS}}

\title{Generalized Parton Distributions and Transversity in
Nucleons and Nuclei.}

\author{S. Liuti \address[MCSD]{Physics Department, 
        University of Virginia, \\ 
        382 McCormick Road, Charlottesville, VA 22904, U.S.A.}%
        \thanks{This work is supported by the U.S. Department
of Energy grant no. DE-FG02-01ER41200. 
} }

\begin{document}

\maketitle

\begin{abstract}

\end{abstract}

\section{INTRODUCTION}
Since they were first introduced in the mid '90s \cite{DMul,Ji1,Rad}
Generalized Partons Distributions (GPDs) have sensibly transformed 
our view of hadronic structure. In fact, in addition to providing a framework
to describe in a partonic language the orbital angular momentum carried 
by the nucleon's constituents, they also give direct new information on the 
partonic distribution in the transverse direction with respect to the 
large longitudinal momentum in the reaction. In \cite{Bur,Diehl03}
GPDs were shown to be related by 
Fourier transformation to the Impact Parameter dependent 
Parton Distribution Functions (IPPDF), originally defined by Soper 
\cite{Soper}. 
However, GPDs are not directly related to the so-called
Unintegrated Parton Distributions (UPDs) appearing {\it e.g.} in 
transverse spin polarized reactions, since the transverse coordinate 
characterizing both GPDs and IPPDFs, is not Fourier 
conjugate to the intrinsic transverse momentum in a UPD. 
A relation can instead be established between the UPDs and 
the non diagonal elements of the GPD (IPPDF) matrix \cite{LiuTan1}.    

All of the recent studies connecting  
coordinate space and momentum space descriptions 
promise a whole new dimension for 
studying hadronic structure that has just begun to 
unravel: GPDs and UPDs are in fact themselves projections
of a more comprehensive theoretical quantity describing a 
seven-dimensional phase space, known as Wigner Distribution
(WD) \cite{BelJiYuan}. 

A number of new efforts to establish a phenomenology that would 
allow one to interpret experimental measurements in terms of GPDs, 
UPDs, {\it ... etc.}, exist (for a review see {\it e.g.} \cite{Diehl03}).
In this paper, in particular, we explore the role of GPDs in nuclei, 
both as a tool to access hadronic configurations' radii 
(cf. the Color Transparency (CT) hypothesis \cite{RalPir,LiuTan1,BurMil}),
and as a method to study the nature of nuclear medium induced 
modifications of the quark and gluon structure of hadrons.    

\section{DEFINITIONS}
 
\begin{figure}
\begin{minipage}{16cm}
\hspace{3cm}
\includegraphics[width=10.cm]{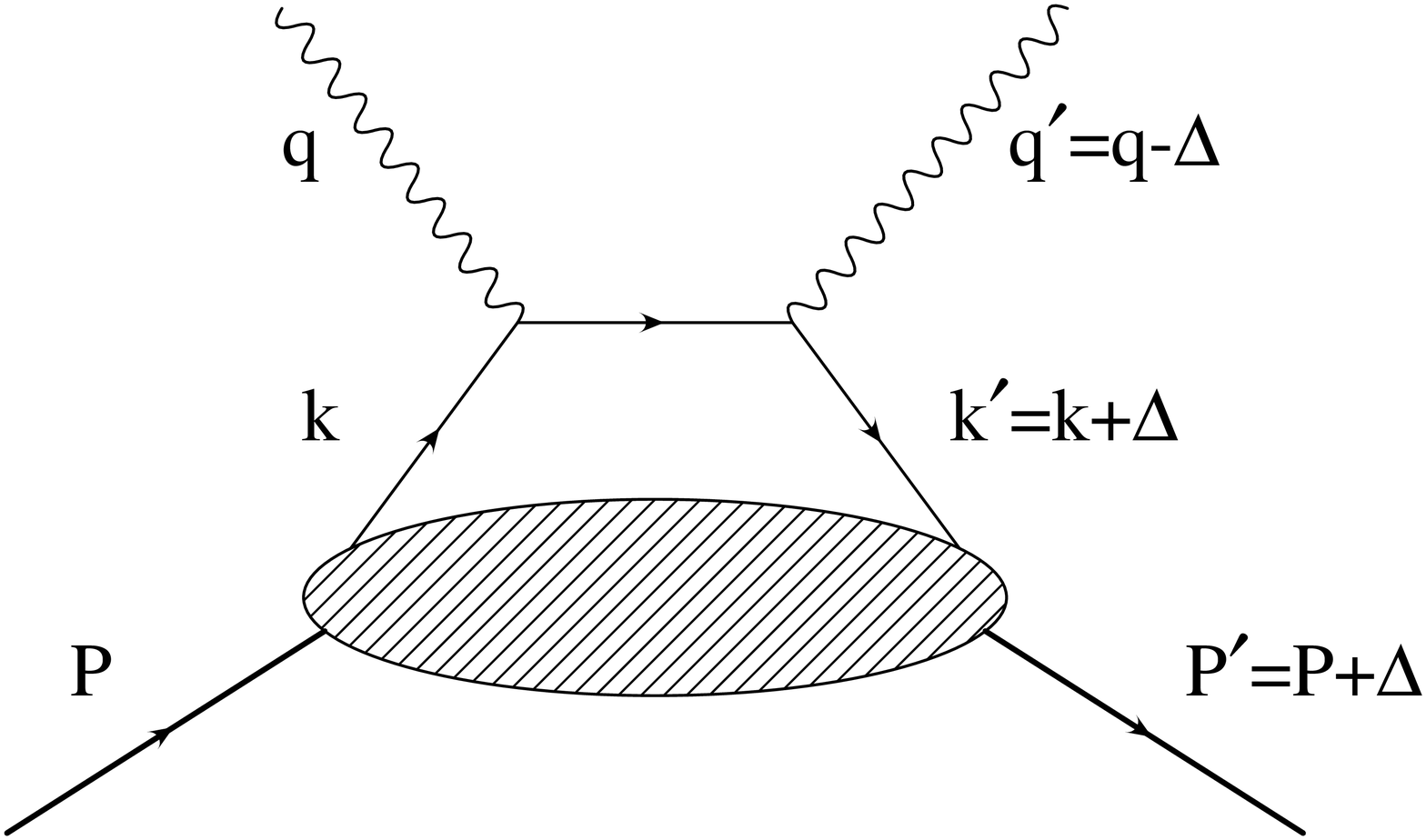}
\caption{Amplitude for deeply virtual Compton scattering process 
at leading order in $Q^2$}
\label{fig1}
\end{minipage}
\end{figure}

GPDs are most easily defined as the structure functions 
that appear in the deeply virtual Compton scattering 
reaction $ep \rightarrow e^\prime p \gamma$ depicted in Fig.\ref{fig1}.
Two GPDs denoted by $H$ and $E$, 
corresponding to the two possibilities for the 
final particle's helicity, describe the process. 
The kinematical invariants that $H$ and $E$ depend on are 
defined respectively to a  
``reference vector''. This can be either
$\overline{P}=(P+P^\prime)/2$, the average nucleon momentum, or $P$, 
the initial nucleon's momentum. In the first case, one has:
$x = (k+ k^\prime)^+/(P+P^\prime)^+$, $\xi=-\Delta^+/2 \overline{P}^+$, and
$t= -\Delta^2_\mu$ \cite{Ji1}; 
in the second, $X = k^+/P^+$, $\zeta = - \Delta^+/P^+$, and $t-\Delta^2_\mu$
\cite{Rad}. The two sets of variables are completely equivalent. 
The set $(X, \zeta, t)$ is, however, best suited both for considering 
the convolution
with nuclear variables, as well as as for perturbative QCD evolution.

A relationship was found by Burkardt \cite{Bur} between
GPDs and the IPPDFs -- the joint distribution,
$\displaystyle d n/dx d{\bf b} \equiv q(x,{\bf b})$ representing
the number of partons of type $q$ 
with momentum fraction $x=k^+/P^+$, located at
a transverse distance ${\bf b}$ (${\bf b}$ is the impact parameter) 
from the center of $P^+$ of the system
\cite{Soper}: 
\begin{eqnarray}
& q(x,{\bf b})  = & \int \frac{d^2 {\bf \Delta}}{(2 \pi)^2} \,
e^{-i {\bf b} \cdot {\bf \Delta}} H_q(x,0,-{\bf \Delta}^2)
\label{bdis1}  \\
& H_q(x,0,-{\bf \Delta}^2)  = & \int d^2 {\bf b} \,
e^{i {\bf b} \cdot {\bf \Delta}} q(x,{\bf b}) .
\label{bdis2}  
\end{eqnarray}
Since $q(x,{\bf b})$ satisfies positivity constraints and it can
be interpreted as a probability distribution, $H_q(x,0,-{\bf \Delta}^2)$
is also interpreted as a probability distribution, namely the Fourier 
transformed joint probability distribution 
of finding a parton $q$ in the proton with longitudinal momentum fraction 
$x$, at the transverse position ${\bf b}$, with respect to the
center of momentum of the nucleon.
The {\em radius} of the system of partons, 
which is needed for quantitative CT studies, is \cite{LiuTan1}:
\begin{equation}
\displaystyle \, \langle {\bf r}^2(x) \, \rangle ^{1/2} = 
MAX\left\{ \langle {\bf b}^2(x) \rangle ^{1/2}, \langle {\bf b}^2(x) \rangle ^{1/2}
\frac{x}{1-x}   \right\}.
\label{r}
\end{equation} 
The UPD, $f(x,{\bf k})$, is defined as \cite{LiuTan1}: 
\begin{equation}   
f(x,{\bf k})  =  \int d^2 {\bf b} \int d^2 {\bf b}^\prime \, 
e^{i {\bf k} \cdot ({\bf b} - {\bf b}^\prime)} \, q(x,{\bf b}, {\bf b}^\prime),
\label{fk}
\end{equation}
where $q(x,{\bf b}, {\bf b}^\prime)$ is the non-diagonal IPPDF, namely
$q(x,{\bf b}, {\bf b}^\prime) \rightarrow q(x,{\bf b})$ for 
${\bf b} \rightarrow  {\bf b}^\prime$.

\begin{figure}
\includegraphics[width=5.2cm]{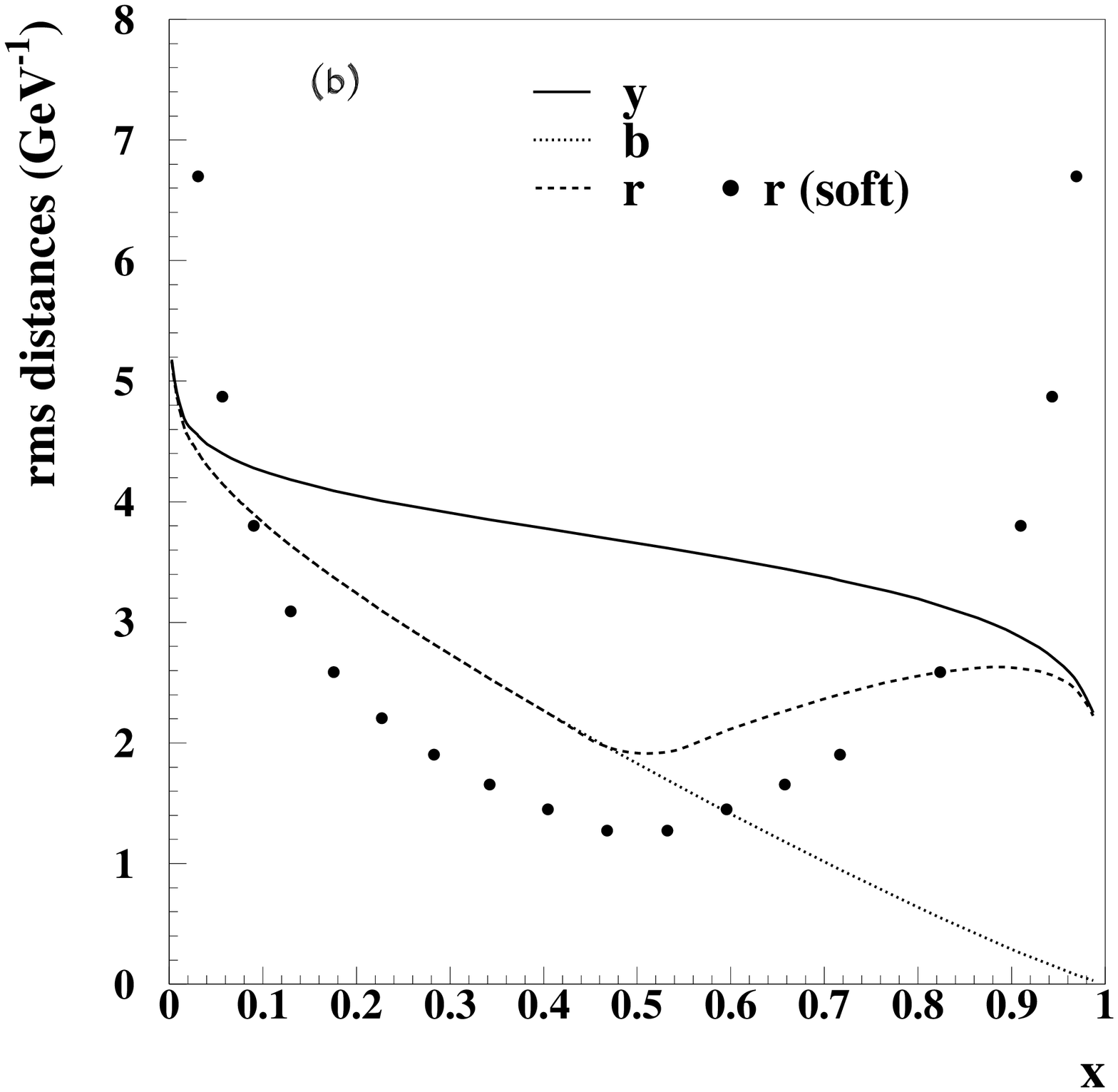}
\includegraphics[width=5.2cm]{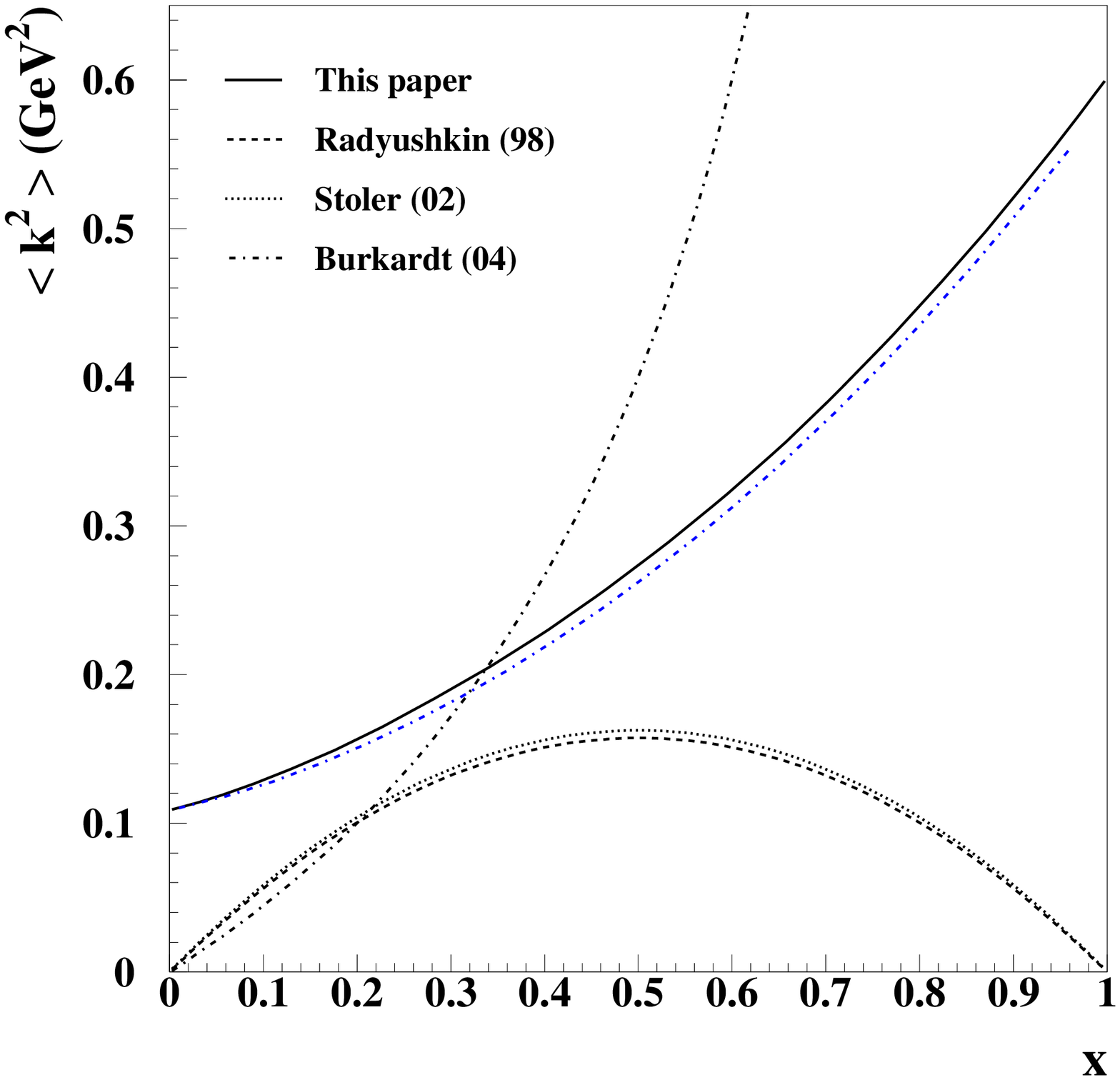}
\includegraphics[width=5.2cm]{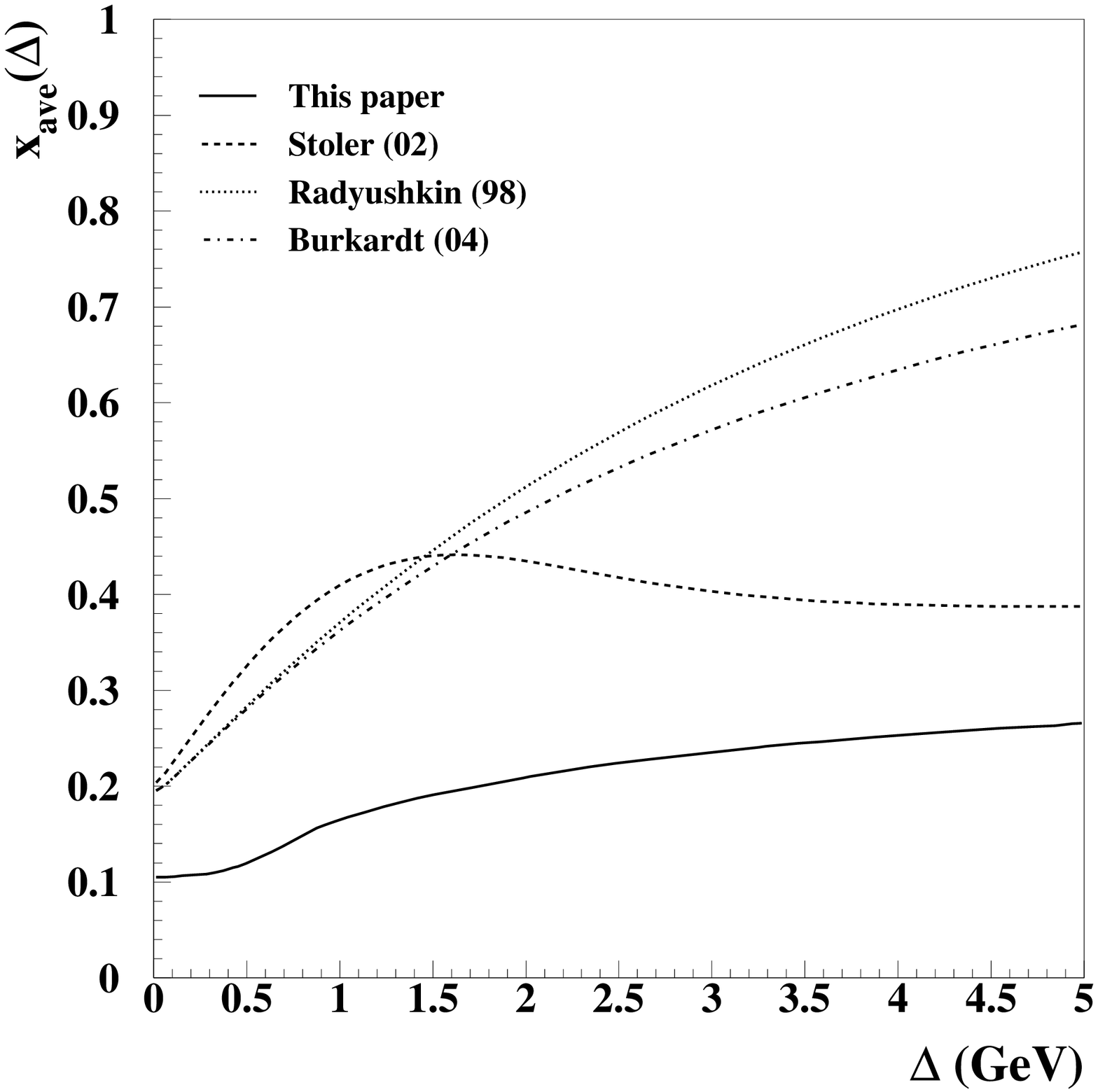}
\caption{The hadronic configuration's radius, Eqs.(\ref{r}) and (\ref{b}), (left);
the intrinsic transverse momentum, Eqs.(\ref{k}) and (\ref{fk}), (center);
and the average value of $x$, as a function of 
$\Delta=\sqrt{-t}$, Eq.(\ref{x}) (right) (adapted from \protect\cite{LiuTan1}).
\label{fig2}}
\end{figure}

Despite $q(x,{\bf b})$ is not directly related to $f(x,{\bf k})$, 
${\bf b}$ and ${\bf k}$ not being 
Fourier conjugates of one another, one can describe the behavior of: 
{\it i)} $\langle {\bf r}^2(x) \rangle^{1/2}$,
which is written in term of, $\langle {\bf b}^2(x) \rangle^{1/2}$;
{\it ii)} the average intrinsic transverse momentum, $\langle {\bf k}(x) \rangle^{1/2}$;
{\it iii)} 
the average value of $x$,  $x(\Delta)$,
by using a consistent set of nucleon vertex functions to model the soft part
of Fig.\ref{fig1}. One obtains:  
\begin{eqnarray}
\displaystyle
\langle \, {\bf b}^2(x) \, \rangle & = & {\cal N}_b \int d^2 {\bf b} \; q(x,{\bf b})
\, {\bf b}^2, \label{b}
\\
\langle \, {\bf k}^2(x) \, \rangle & = & {\cal N}_k \int d^2 {\bf k} \; f(x,{\bf k})
\, {\bf k}^2,
\label{k}
\\
\langle x(\Delta) \rangle & = & {\cal N}_x \int_0^1 dx \, x \, H(x,\Delta)
\label{x}
\end{eqnarray}
where ${\cal N}_b$, ${\cal N}_k$ and ${\cal N}_x$ are normalization factors, and
$\Delta=\sqrt{-t}$.
The quantities in Eqs.(\ref{b})-(\ref{x}) are displayed in Fig.\ref{fig2}.

\section{PROBING THE TRANSVERSE STRUCTURE OF BOUND NUCLEONS}
Nuclei have since long been suggested as ``laboratories'' to observe
several aspects of quarks and gluons dynamics.  
One can for instance detect small size hadronic configurations 
by studying the passage of hadrons through nuclear matter, 
and the conditions for the onset of Color
Transparency in exclusive reactions of the type 
$eA \rightarrow e^\prime p(A-1)$, 
or $p (\pi) A \rightarrow p^\prime (\pi^\prime) \, p (A-1)$, 
$\gamma A \rightarrow \pi N (A-1)$. 
One can also learn about modifications of the confinement size of nucleons
embedded in the nuclear medium through 
inclusive deep inelastic experiments (the so-called EMC effect). 

Both exclusive and inclusive types of reactions involve in their description 
transverse degrees of freedom of the hadrons. 
In what follows we briefly describe this particular aspect, 
and we show with a few examples, the prominent role and new insight provided
by GPDs.     

\subsection{Color Transparency}
In the hard scattering approach to QCD,
exclusive reactions, and similarly inclusive reactions 
at $x_{Bj} \approx 1$ (where $x_{Bj} =Q^2/2M\nu$, $Q^2$ and $\nu$ being 
the four-momentum transfer squared and the energy transfer, respectively) 
are expected to be
dominated by hadronic configurations with the 
minimum number of quarks (anti-quarks), located 
within a small relative transverse 
distance, $ \approx 1/\sqrt{Q^2}$, 
(see {\it e.g.} \cite{Ji_recent} and references therein).
However, lacking any direct experimental proof, 
it is also possible to envisage situations where the transverse size
of exclusive hard processes might not be small, as {\it e.g.}
in \cite{Hoyer}, due to the persistence
of large endpoint contributions to the hadron's wave 
function.
 
Small distances can in principle be filtered by studying either the $Q^2$ 
dependence of nuclear cross sections, or the dependence on the atomic
number, $A$, at finite (moderate) $Q^2$. Large separations 
are in fact expected to gradually be blocked
by the strong interactions occurring in the nucleus 
since the cross section for hadronic
rescatterings  
is proportional to the hadrons transverse size, as dictated by
the Chew-Low-Nussinov mechanism (see \cite{RalPir} and references 
therein).
From a practical point of view, however,  
current searches for CT
might appear to be in a stall as 
all experiments performed so far seem not to show on one side, 
any marked trend for the onset of 
this phenomenon, and, on the other, the observables do not allow one to
discern what factors ({\it i.e.} features of the hadronic interactions, 
or of the hadronic wave function, or else...) 
are responsible for any lack of CT. 

Generalized Parton Distributions (GPDs) seem to provide 
the best candidates to explore the 
existence and observability of small size hadronic 
configurations.
For illustration, in Ref.\cite{LiuTan1} we 
considered the $(e, e^\prime p)$ process from a nuclear target
where we introduced a nuclear filter for the 
large transverse size components as follows:
\begin{equation}
\Pi(b) = \left \{  \begin{array}{c} 
1  \; \; b <    b_{max}(A) \\
0 \; \;  b \geq b_{max}(A) \end{array} \right. ,
\end{equation}
$b_{max}(A)$ being the size of the filter. 
The transparency ratio is then defined as: 
\begin{eqnarray}
\displaystyle T_A(Q^2)  =   
\frac{\left[ \int_0^1 dx H_A(x,t) \right] ^2}{\left[ \int_0^1 dx H(x,t) \right]^2}  
= \frac{\left[ \int_0^1 dx \int_0^{b_{max}(A)} db \, b \, q(x,b) 
J_0(b\sqrt{t}) \right] ^2}{\left[ \int_0^1 dx H(x,t) \right]^2}  
\end{eqnarray}
By varying the parameter $b_{max}$, and by using   
different parametrizations of $q(x,b)$, one can in principle disentangle
the effect of the hadronic size from the effect of the hadronic
interactions in the nuclear medium.   


\subsection{Nuclear Deep Inelastic Scattering}
GPDs provide also a unique tool to describe 
the spatial distribution of quarks and gluons in nuclei. 
Throughout the years since the first discovery of the EMC effect
\cite{EMC}, an increasingly coherent picture has emerged of
Deep Inelastic Scattering (DIS) processes from nuclei.
The main outcome is that nucleons, despite the high resolution
achieved in DIS experiment, do not behave 
as free. Their interactions 
are instead important, and they are responsible for the modifications of 
the nuclear cross section with respect to the free nucleon one.
Despite the general consensus on this picture,  
the way these interactions proceed is still largely model dependent, 
ranging from increasingly sophisticated binding models \cite{AKL}, 
effective theories \cite{Mil_A,Tho_A}, and
``rescaling'' of the scale dependence of the effect.  
In this paper we use an approach where we account for 
final state interactions between the outgoing nucleon and nuclear debris, 
parametrized as off-shell effects. 
\footnote{The four-momentum squared of a nucleon inside the nucleus
is different from its mass squared, $P^2 \neq M^2$, and
it is instead related to the nucleons transverse momentum, $P_T$.} 
An important aspect of our approach is that 
it provides a description of the EMC effect 
that, at variance 
with the naive (on-mass-shell) binding models, can simultaneously 
reproduce both the $x_{Bj}$ and $A$ dependences of the data. 
We present results for a spin 0 nucleus, namely $^4He$ (more details
and evaluations for larger nuclei can be found in \cite{LiuTan2}).
The GPD, $H_A$, reads: 
\begin{equation}
H^{A}(X,0,t)= \int \frac{d^2 P_{\perp} dZ}{2(2\pi)^3} \, \rho_{A}(P,P^{\prime}) \, H^{off, N}(X_N,0,P^2,t),
\label{HA}
\end{equation}
where we used the $(X,\zeta,t)$ set of variables. Moreover, in a nucleus 
$X  = k^+/(P_A^+/A)$, $Z  =  P^+/(P_A^+/A)$, 
and $X_N  =  X/Z \equiv  k^+/P^+$, $k$, $P$, $P_A$ being the active quark,
nucleon, and nuclear momentum, respectively.  
For an 
off-shell nucleon, and for $\zeta =0$, $H^{off, N}$ is defined as:
\begin{eqnarray}
H^{off, N}(X_N,0,P^2,t)  =  \frac{X_N}{1-X_N}
\int \frac{d k_\perp^2}{2 \pi} \,
\rho_{N}(k(P), k^{\prime}(P)),
\label{bound_F}
\end{eqnarray}
$\rho_{A}(P,P^{\prime})$ and $\rho_{N}(k(P), k^{\prime}(P))$ 
are {\it off-diagonal} 
nuclear and nucleon spectral functions,
respectively.
Notice that $H^{off, N}(X_N,0,P^2,t)$ is  
modified both kinematically and dynamically with respect to
the free nucleon GPD. 
Kinematical modifications 
due to Fermi motion and nuclear binding produce a shift
in the $X$ dependence with respect to the free nucleon. 
$H^{off, N}(X_N,0,P^2,t)$ is however also 
structurally different from the on-shell case.   

Off-shell modifications, differently from Fermi motion and binding,
affect the transverse variables. 
It is therefore of the outmost 
importance to evaluate carefully their impact on GPDs,
especially in view of the fact that these are 
Fourier transforms of IPPDFs. GPDs in fact, provide a handle to 
directly evaluate the spatial modifications of the nucleon inside
the nuclear medium. By making the assumption that the struck nucleon
is located at the center of the nucleus, {\it i.e.} the nucleon
impact parameter ($\beta$) distribution ia described by: 
$\widetilde{\rho}_A(Z,\beta) \approx \rho_A(Z,P_\perp) \times \delta(\beta)$, we obtain 
for a bound nucleon
\begin{eqnarray}
\langle b_N^2(x) \rangle_{Bound}  =  
\frac{\int d^2{\bf P}_\perp \int_X^A 
dZ \; \left[ \int d^2 {\bf b} \, q(X/Z,{\bf b}) \, {\bf b}^2 \right] \rho_A(Z,P_\perp^2) }{\int d^2
{\bf P}_\perp \int_X^A dZ  \; f_N^{OFF}(X/Z,P_\perp^2) \rho_A(Z,P^2)},
\label{bbound}
\end{eqnarray}
where $f_N^{OFF}(X/Z,P_\perp^2)$ is the PDF in an off-shell nucleon \cite{AKL}.

In Fig. 3 we present results for: {\it i)} the average impact parameter squared 
in a bound nucleon, 
Eq.(\ref{bbound}); {\it ii)} the intrinsic transverse momentum in a nucleus, calculated
by convoluting Eq.(\ref{k}) with $\rho_A(Z,P_\perp^2)$; {\it iii)} the ratio 
$\displaystyle R=[H_A(X,t)/F^A(t)]/[H_N(X,t)/F_1(t)]$, $F_A$ being the form factor for $^4He$
and $F_1$ being the integral of Eq.(\ref{bdis2}). All results are part of a preliminary 
study accounting for the effect of binding and Fermi motion. We find that the main nuclear 
effect on $\langle b^2 \rangle$ is an enhancement at large $x$ due to Fermi motion. Similarly,
the average intrinsic $k_\perp$ is enhanced, as a function of $x$ due to the combined effect
of both longitudinal and transverse motion inside the nucleus. Finally, the impact of nuclear
effects on GPDs is best studied by normalizing $H_{A,N}$ to the corresponding form factors. 
We find that both the effect of binding (dip at intermediate $X$), and of Fermi motion
are enhanced at $t \neq 0$. The effects of Fermi motion are in fact sizable at $x \approx 0.6$,
a region more accessible experimentally.    

\begin{figure}
\includegraphics[width=5.2cm]{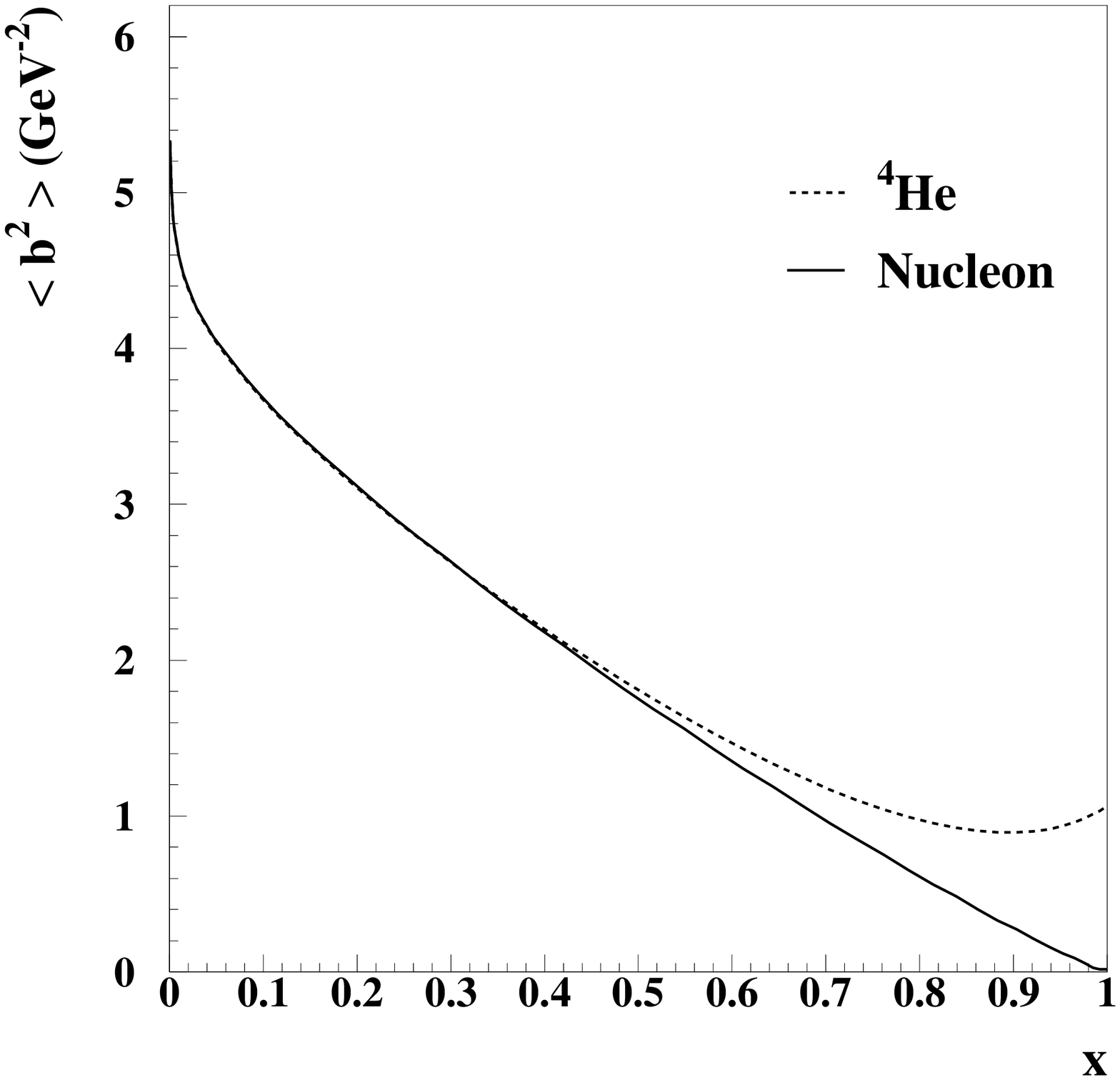}
\includegraphics[width=5.2cm]{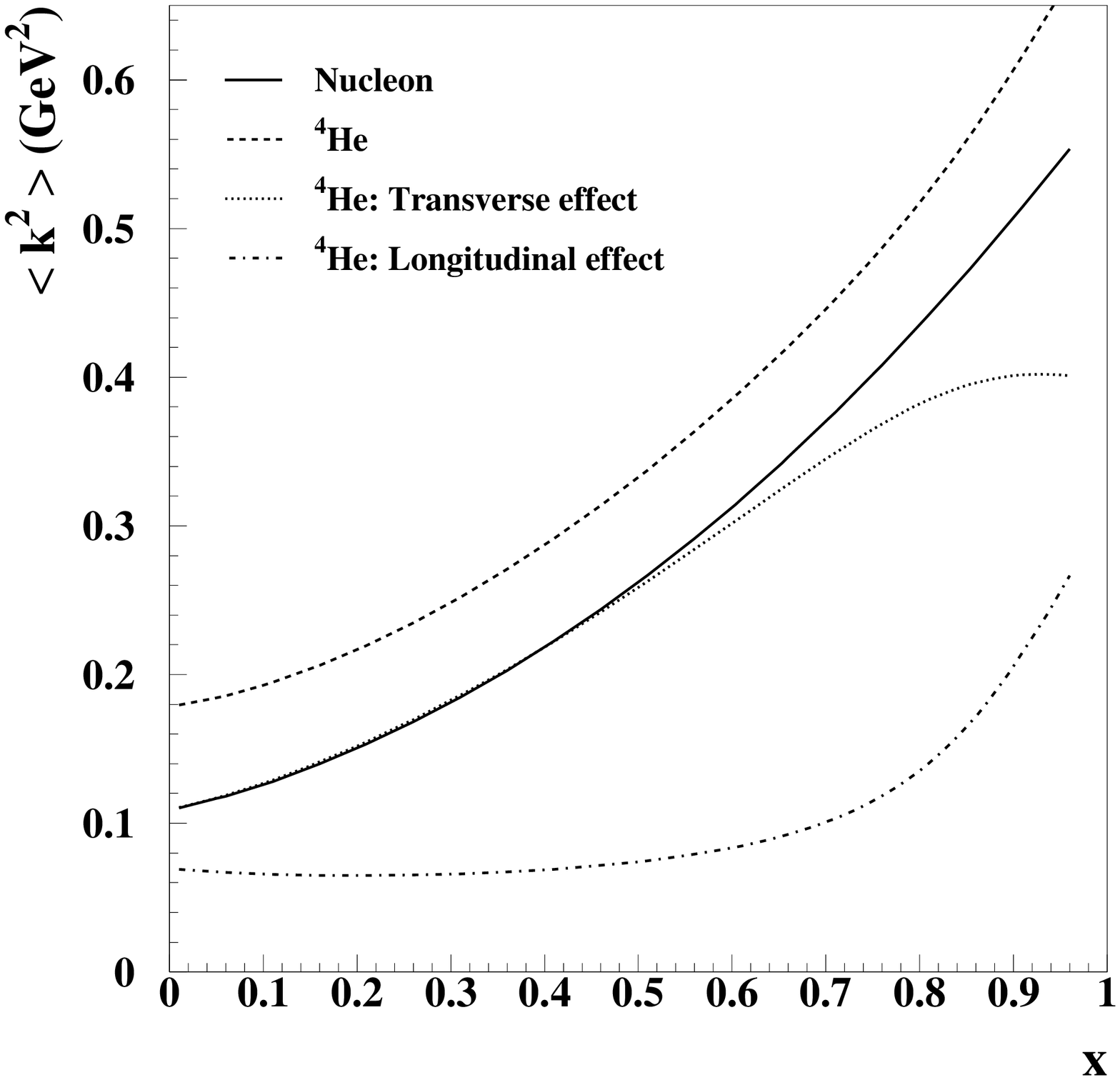}
\includegraphics[width=5.2cm]{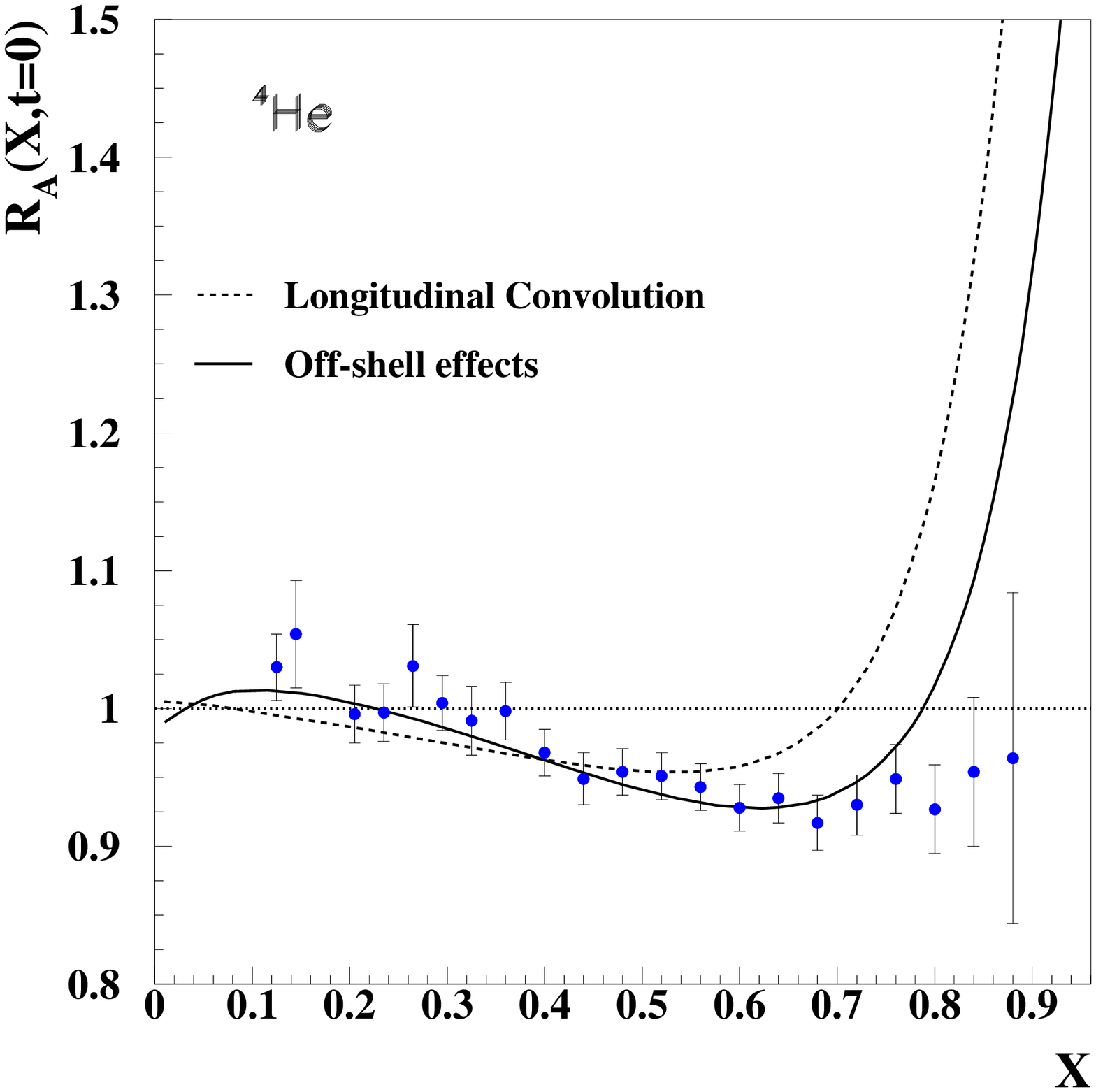}
\caption{The quark's radius in a nucleon inside $^4$He (left);
the average intrinsic transverse momentum in a nucleon inside $^4$He (center);
and the ratio $R=[H_A(X,t)/F^A(t)]/[H_N(X,t)/F_1(t)]$ of GPDs in $^4$He to
the free nucleon one, normalized by the corresponding form factors (right).
\label{fig3}}
\end{figure}

In conclusion our approach for studying nuclear effects in both hard exclusive 
and inclusive processes using GPDs 
will allow for a more detailed understanding of the ``intrinsic'' transverse 
components in nuclei. In particular, it will be possible to determine whether nucleons'
deformations in the nuclear medium are at the
origin of EMC effect.

\end{document}